\documentclass[sn-mathphys]{sn-jnl}
\usepackage[super,compress]{cite}
\usepackage{graphicx}
\begin{document}

%
%

\title{Dissociation rates and recombination likelihood of bottomonium states in heavy-ion collisions at $\sqrt{s_{\rm NN}}$ = 5.02 TeV}

\author*[1]{ \sur{Abdulla Abdulsalam}}\email{akizhakkepura@kau.edu.sa}

\author[2]{ \sur{Mohsin Ilahi}}\nomail

\author[1]{ \sur{Lana alabbasi}}\nomail

\author[3]{ \sur{Rafeeda Kassim}}\nomail

\affil[1]{\orgdiv{Department of Physics}, \orgname{King Abdulaziz University}, \orgaddress{\city{Jeddah}, \postcode{21589}, \country{Saudi Arabia}}}

\affil[2]{\orgdiv{Department of Physics}, \orgname{Aligarh Muslim University}, \orgaddress{\city{Aligarh}, \postcode{202002}, \country{India}}}

\affil[3]{\orgdiv{Physical Science Division}, \orgname{Puthiyangadi Jama-ath Higher Secondary School}, \orgaddress{\city{Madayi}, \postcode{670304}, \country{India}}}


\abstract{In the medium of relativistic heavy-ion collisions, dissociation of the quarkonium and their survival
  have been studied to understand the properties of Quark Gluon Plasma (QGP). The coupled rates of dissociation
  and recombination reactions in QGP are commonly solved with the Boltzmann transport equation in which the formation
  and dissociation reactions compete with each other. Since the dissociation of newly formed bound-states is not
  accounted in the Boltzmann equation, a framework of decoupled rates is developed to assess the combined effect
  of gluon-induced dissociation and recombination (though it is small for $\Upsilon$) together with color screening
  on bottomonium production in heavy-ion collisions at center of mass energy ($\sqrt{s_{\rm NN}}$) = 5.02 TeV.
  To calculate the recombination rates, we have employed an effective method of Bateman solution which makes
  sure the correlated effect between the recombination and the dissociation of the newly combined bottomonium
  in the QGP medium. The modifications of bottomonium have been estimated in an inflating QGP with the
  constraints matching with the dynamics of Pb+Pb collision events at LHC.}

\keywords{Quark Gluon Plasma, Bottomonium, Recombination, Dissociation, Decoupling rate equations.}

\maketitle


\section{Introduction}\label{Intro}

The Quark Gluon Plasma (QGP) and its properties have been studied for more than three decades.
With the help of relativistic heavy ion collisions (HIC), comprehensive studies are being performed on the strongly interacting
nuclear matter at high energy density ($>$ 1 GeV/fm$^3$) and temperature ($>$ 500 MeV). A phase transition to the
deconfined matter of quarks and gluons is predicted when the temperature of the nuclear matter is above
a critical temperature~$T_C \sim$ (150-175) MeV~\cite{QGP_Tc}. The modification of quarkonium yields and
their suppression in the QGP medium have been assessed for many years as the signature of the QGP formation. 
Main reasons of the suppression can be Debye color screening~\cite{SATZ}, gluon-dissociation, cold
nuclear matter effect (CNM) and so on. Earlier experiments at RHIC and LHC (CMS, ATLAS, and ALICE) have performed
measurements of charmonia and bottomonia at different energies~\cite{Upsi_CMS_NPA,ALICE502TeV_Psi,ATLAS502,JCMS502TeV,JCMS502TeV_Abd}.
These measurements support the hypothesis of enhanced suppression of less strongly bound states. 
It has been observed that out of  the three bottomonium states~\cite{CMS502TeV_UpsiRaa}
the $\Upsilon(3S)$ is the most suppressed whereas the $\Upsilon(1S)$ is the least suppressed one.
All measurements are indicating the predicted sequential suppression of the quarkonia. This gives hints
of the formation of a high temperature QGP medium. Some of the experiments are throwing hints of
recombination/regeneration of quarkonia~\cite{ALICE502TeV_Psi,ALICE276TeV_Psi}. The regeneration for the bottomonium,
due to its heavier mass, is less expected but can be nontrivial at higher energy collisions ($>=$ 5.02 TeV)
~\cite{CoupledBoltz,Upsi_Reco}. Although several studies have been carried out in various directions,
a thorough investigation is still required to parameterize the different medium-induced reaction effects on the states
apart from the modification due to non-QGP reactions, such as nuclear absorption, multi-particle interactions,
etc.~\cite{Vogt,JpsiRatioNPA,DURapp}. Also, the determination of initial parameter values and their variation
has a remarkable impact on the modification of the quarkonium yields. 
The robustness of the recombination mechanism in kinetic and statistical
models greatly depends on the hadronization of the quarkonium states and thermal equilibration of the charm/bottom
quarks in the medium~\cite{Andron_Charm,IslamStrickland}.

In kinetic models, the time evolution of quarkonium states in the color deconfinement medium are composed by
the Boltzmann transport equation which describes the coupled rates of dissociation and recombination~\cite{Stachel_Reco,Thews_Mangano}.
As collisions energy is increasing, the probability of the recombination goes higher simultaneously
with dissociation rate of newly formed bound states. Unlike in the solution of Boltzmann equation, this
additional decline of the bottomonium, the theme of this paper, is to be accounted to assess the total survival
rates during the QGP lifetime. In this study, we developed a framework of decoupled rate equations of dissociation
and recombination for bottomonium along with an extended color screening model~\cite{URatioAbd} and calculated the total
survival probabilities of bottomonium states($\Upsilon(nS)$)
in the deconfined medium of QGP. The decoupled recombination equation is solved with the help of Bateman
equation of chain decay reactions in order to evaluate the dissociation of newly formed bottomonia~\cite{Bateman}. Also it's
good approach adoptable to quantify the different effects.
The competing factors such as the formation time $\tau_{F}$, medium temperature $T(\tau)$, lifetime $\tau_{QGP}$ and dissociation temperature  $T_D$ determine the pattern of the survival rates of $\Upsilon(nS)$ states in the
kinematics of transverse momentum, $p_{T}$ and centrality, $N_{part}$ (geometry of the HIC). Details of similar kind of the other suppression model are available in Ref.~\cite{ScriptaAbd, SuppVineetShukla,Fransua}.

\section{Theoretical framework}
\subsection{Initial production of Bottomonium}

The initial production of the bottomonium states and their yields in each centrality of the collision are
important inputs for the model calculations. The production of primordial yields $N_{b\overline{b}}$ and the
processes like hadronization, nuclear absorption and shadowing effects determine the final number of
yields, $N_{\Upsilon}$ = $\sigma_{dir} \cdot L \cdot B.R$. Here $\sigma_{dir}$ is the direct cross-section per
nucleon pair, $B.R$ is the branching fraction of $\Upsilon$(nS) decay into two dimuons~\cite{Vogt, PDGReview} and
$L$ (= 368 $\mu b^{-1}$) is the collision luminosity in Pb+Pb collisions at $\sqrt{s_{\rm NN}}$ = 5.02 TeV.
The corresponding yields are multiplied by an exponential function to get proper spectrum of transverse momentum $p_T$.
The $p_T$ spectrum of the excited states are scaled by the ratios of $\Upsilon$(2S)/$\Upsilon$(1S) and
$\Upsilon$(3S)/$\Upsilon$(1S) respectively measured in pp collisions at $\sqrt{s}$ = 2.76 TeV~\cite{CMSU2,CMS7TeVpp_Upsi1s2s3s}.
The important parameters of the model like $T_C$, initial formation time of QGP $\tau_{0}$, $\tau_F$, $T_D$, etc.
are obtained either from temperature-dependent potential models or from Lattice QCD calculation~\cite{QGP_Tc,Initials1}.
The variation in these inputs can trigger effective changes in the final results of dissociation and recombination.
The values of cross-sections and branching ratios are given in Table~\ref{SigmaDir}.

\begin{table}[ph]
\begin{center}
\caption{The direct cross-sections per nucleon pair for bottomonium states in Pb+Pb collisions at $\sqrt{s_{\rm NN}}$ = 5.02 TeV ($\sigma_{dir}$/nucleon pair($\mu$b)) and the branching fraction of $\Upsilon$(nS) decay into two dimuons~\cite{Vogt,PDGReview}.} 
\begin{tabular}{@{}cccccc@{}}
                 \hline 
			{\rm } & $\Upsilon$(1S) &$\Upsilon$(2S) &$\Upsilon$(3S) &$\chi_b$(1P) &$\chi_b$(2P)  \\
			\hline \\
                            {\rm Pb+Pb = 5.02 TeV} &0.15 &0.092 &0.055 &0.03 &0.023 \\
                            \\
                            {\rm B.R($\%$) $\rightarrow \mu^{+}\mu^{-}$} &2.482 &1.93 &2.18 &0.48 &0.266 \\
			\hline
		\end{tabular}\label{SigmaDir}
\end{center}
\end{table}

\subsection{Debye color screening}

According to the model, the QGP formed at an initial time $\tau_{0}$ with entropy density $s_0$ and temperature $T_0$ 
undergoes an isotropic expansion. The basic formalism of the model is well described in~\cite{JpsiRatioNPA,URatioAbd}.
A bottom-quark pair can escape the color-charge screening region $r_D$ and form a bound state if the position at 
which it is created satisfies

\begin{equation}\label{taumax}
	\| {\rm \bf r} + {\tau_{F} {\rm \bf p_{T}} \over M} \| > r_D,
\end{equation}
where the screening region $r$ $<$ $r_D$ is shrinking because of the cooling of the system. Let $\Theta$ to be
the angle between ${\rm p_{T}}$  and ${\rm \bf r}$, then Eq.~(\ref{taumax}) provides a range of $\Theta$ in which
the bottom-quark pair can escape~\cite{JpsiRatioNPA}.

\begin{equation}\label{cosmax}
	{\rm cos} \, \Theta \ge V   ~~~~{\rm where } ~~~~ \nonumber \\ 
	V = \frac{ r_D^2 - r^2 - (\tau_{F}p_{T}/M)^2}{2r \,(\tau_{F}p_{T}/M)}.
\end{equation}

Now, the probability of a bottom-quark pair ($\sigma(r)$) to be created at position $r$ which is symmetric
in transverse plane, is parameterized as
\begin{equation}
	\sigma(r) = \left(1 - \left( {r \over R} \right)^2\right)^{1/2}.
\end{equation}

Combining with the screening region and the $\sigma(r)$, the survival probability of bottomonia as a function
of screening radius becomes 
\begin{equation}\label{Surv_Color}
	S(R(N_{\rm part})) = \frac{\int_0^Rdr~r~\sigma(r)~\Theta(r,p_{T})}{\pi\int_0^Rdr~r~\sigma(r)} = S_{col}.
\end{equation}

Here $R = R(N_{\rm part})$ is the QGP-medium size (screening region) in terms of the radius of the Pb nucleus
as $R_0 = r_0\, A^{1/3}$. The fireball temperature, $T(N_{\rm part})$ which depends on the collision energy and
size of the nucleus, is calculated in each geometrical centrality of the collisions as in the Ref.~\cite{URatioAbd}, 

\begin{equation}\label{InT1}
T(N_{\rm part})^3 = T_0^3 \, \left({dN/d\eta \over N_{\rm part}/2}\right) / \left({dN/d\eta \over N_{\rm part}/2}\right)_{0-5\%},
\end{equation}
 where $T_0$ is the initial temperature in 0-5\% centrality (most central collision) and $(dN/d\eta)$
is the charge multiplicity as a function of the number of participants measured by ALICE experiment \cite{ALICEmult}. 
With QGP formation time $\tau_{0}$ = 0.15 fm/$c$ at LHC~\cite{Initials1, Initials2}, we obtain the initial
temperature $T_{0}$ in most central collision is 0.65 GeV. The masses and other properties of quarkonium states
which are used in this study are listed in Table~\ref{QuarProp}. 

\begin{table}[h]
  \begin{center}
    \caption{Quarkonium states and their properties~\cite{QPROP,Initials1}.}   
		\begin{tabular}{@{}ccccccccc@{}}
			\hline \\
			{\rm State} & $J/\psi$ &$\chi_{c}$ &$\psi'$ &$\Upsilon$(1S) &$\Upsilon$(2S) &$\Upsilon$(3S) &$\chi_b$(1P) &$\chi_b$(2P)  \\
			\hline \\
			{\rm mass~[GeV]}     &3.10 &3.53 &3.68 &9.46 &10.02 &10.355 &9.86 &10.23 \cr \\
			$\Delta E$ {\rm[GeV]}& 0.64& 0.20& 0.05& 1.10 &0.53& 0.20 & 0.70  &0.34\cr \\
			{\rm r$_0$~[fm]} &0.50 &0.72 &0.90 &0.28 &0.56 &0.78 &0.44 &0.68\cr \\ 
			{\rm T$_D$/T$_c$} &1.4 &1.0 &0.9 &2.0 &1.3 &1.0 &1.3 &1.0\cr \\
                        {\rm $\tau_F$} &0.89 &1.5 &2.0 &0.76 &1.9 &2.0 &2.6 &3.1\cr \\
			\hline \\
		\end{tabular} \label{QuarProp}
	\end{center}
\end{table}

\subsection{Medium dynamics at LHC Experiments}
The relativistic dynamics of HIC is formulated by employing the Bjorken boost invariant scenario exhibiting
accelerated expansion in transverse direction, giving rise to a cylindrical volume geometry related to the collision dynamics.
The volume with an isentropic expansion of the QGP fireball having the dependence of proper time
$\tau$ and radius R$(N_{\rm part})$ is given as  

\begin{equation}
  V(\tau) = \pi R(N_{\rm part},\tau)^{2} (\tau{\bf p_{Z}}/M).
\end{equation}

The term $({\bf p_{Z}}/M)$ is introduced to weigh up the momentum in longitudinal direction($p_{Z}$) related to the quark
pairs and to make sure the longitudinal elongation of the volume during the lifetime of QGP~\cite{JpsiRatioNPA}. With an
acceleration {\bf a} = 0.1 $c^{2}/fm$~\cite{ThermalCharm_Zhang}, the radius in transverse direction becomes,  
\begin{equation}
	R(N_{\rm part},\tau) = R(N_{\rm part}) + a(\tau_{med} - \tau_0)^{2}/2.
\end{equation}
Hence one can get 2-dimensional elongation (longitudinal + transverse) for the volume of the fireball. Also, the
temperature of the fireball volume goes down with respect to the proper time $\tau$ as $T(\tau) = (\tau_{med}/\tau)^{1/3} T_0$.
The $\tau_{med}$ is the time period during which the medium-induced reactions like color screening and gluon-dissociation
are active in each centrality of the collision.

\section{Decoupling dissociation and recombination}

\begin{figure}[htb]
\begin{center}
\begin{tabular}{cc}
 \includegraphics[width=0.49\textwidth]{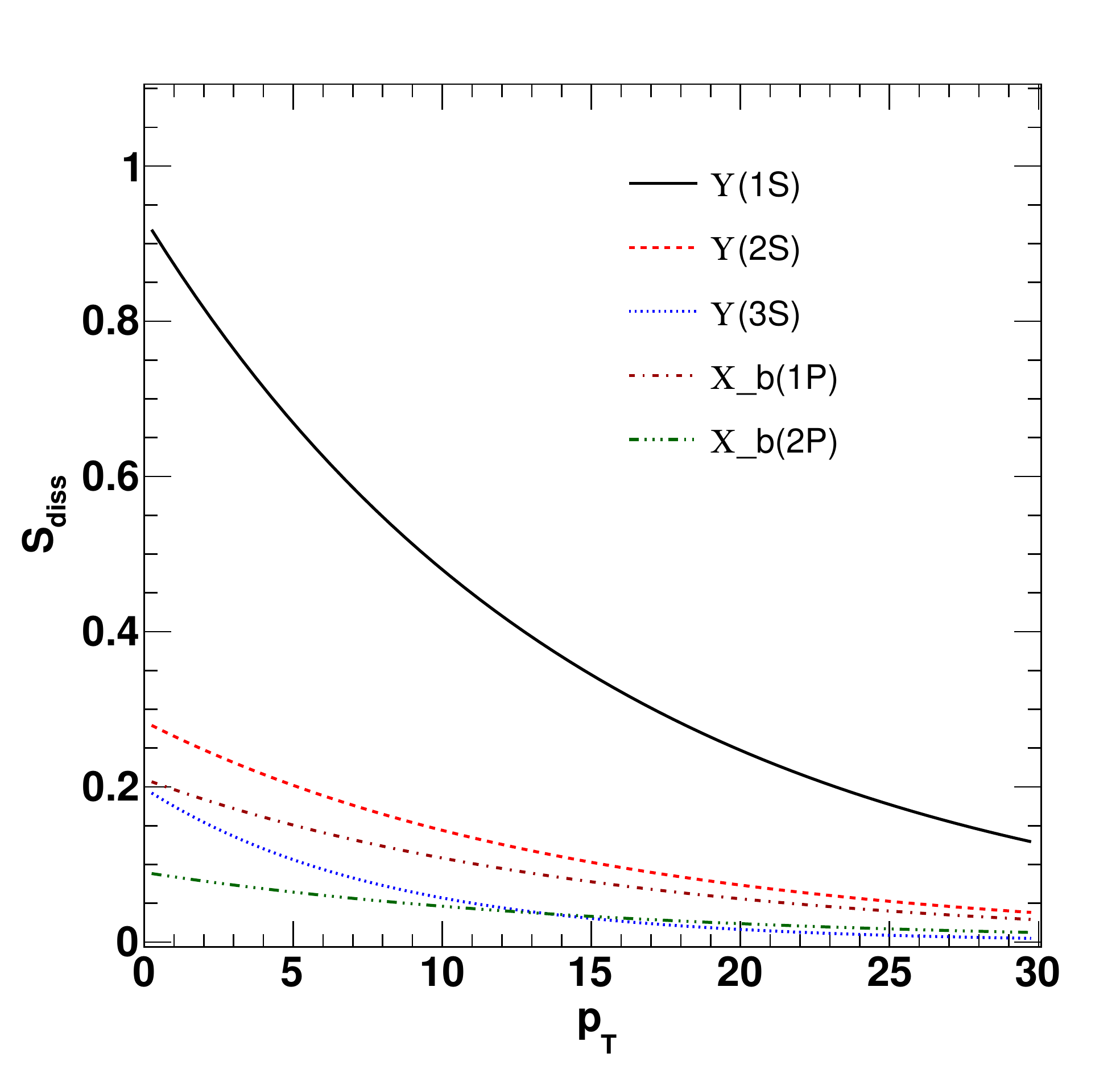}&
\includegraphics[width=0.48\textwidth]{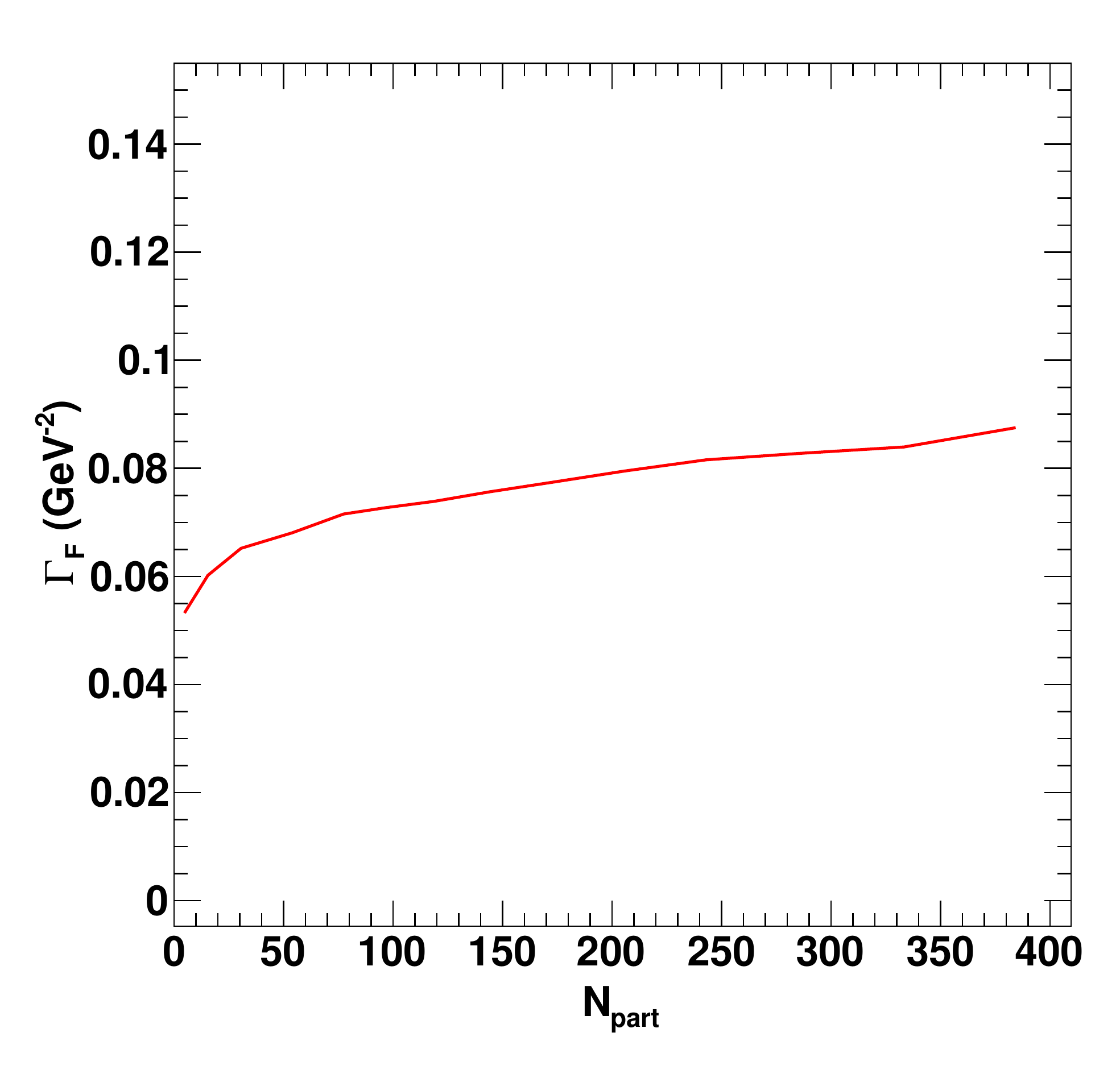}\\
\end{tabular}
\caption{The (Left) shows survival rates of bottomonium states from the gluonic dissociation as a function of p$_T$
  and (Right) shows the formation rate in different centrality regions.}
\label{fig:Diss_Rate}
\end{center}
\end{figure}

The gluonic dissociation of bottomonium and recombination of $b$
and $\overline{b}$ quarks are important processes in the QGP medium.
The time evolution of $b$ quarks and bottomonium states in the deconfined region, according to the Boltzmann
transport equation is

\begin{equation}\label{Bolt_Reco}
\frac{dN_{\Upsilon}}{d\tau} = \Gamma_{F}N_{b}N_{\overline{b}}[V(\tau)]^{-1} - \Gamma_{D}N_{\Upsilon}n_{g} ,
\end{equation}

here $n_g$ represents the number density of gluons which depends on the medium temperature. The widths $\Gamma_{D,F}$
are  dissociation and formation reaction rates $ \langle \sigma v_{rel} \rangle$ respectively averaged over the
momentum distribution of the participants ( $\Upsilon$ and $g$ for $\Gamma_{D}$ and $b$ and $\overline{b}$ for $\Gamma_{F}$ )
~\cite{Stachel_Reco,Re_Prob,CS_Dis}. The recombination rate, although it is less predicted for the bottomonium, an abundant
production of bottom quarks at higher energy collisions ($>=$ 5.02 TeV) makes the rate nontrivial and have to be
calculated for the determination of different reaction effects precisely. Such a recombination rate in Pb+Pb collisions
at $\sqrt{s_{\rm NN}}$ = 5.02 TeV is portrayed in Fig.~\ref{fig:Diss_Rate} (Right). 

The dissociation and recombination are two inverse-competing processes and here, we consider them as two
separate processes and solve corresponding differential equations. 
Then we combine all the rates to get the total probability of survived bottomonium states
(= Nuclear modification factor, $R_{AA}$). The dissociation and the recombination rates are given as 

\begin{equation}\label{Decoup_Dis}
  \frac{dN_{\Upsilon}^{D}}{d\tau} = - \Gamma_{D}N_{\Upsilon}(0)~n_{g}   
\end{equation}

\begin{equation}\label{Decoup_Reco1}
\frac{dN_{\Upsilon}^{F}}{d\tau} = \Gamma_{F}N_{b}N_{\overline{b}}[V(\tau)]^{-1}. 
\end{equation}

For the gluon dissociation rate, the solution of Eq.~(\ref{Decoup_Dis}) provides the number of
bottomonium states that survived the reaction. 

\begin{equation}\label{Decoup_Dis2}
  N_{\Upsilon}^{D} = N_{\Upsilon}(0)~exp^{-\int_{\tau_{0}}^{\tau_{QGP}} \Gamma_{D}n_{g}d\tau}.
\end{equation}

The $N_{\Upsilon}(0)$ (= $\sigma_{\Upsilon}^{NN}~T_{AA}(\tau_0, b)$) represents the number of initially produced bottomonia
in the collisions. $\sigma_{\Upsilon}^{NN}$(= 0.16 mb) is the production cross-section in p+p collision and $T_{AA}$ represents
nuclear overlap function extracted from the Ref.~\cite{Upsi_Grandchamp,Enterria}.
$\Gamma_{D}$, the thermal gluon-$\Upsilon$ dissociation reaction rate is defined as

\begin{equation}\label{Dis_CS_Avg}
 \Gamma_{D} =  \langle v_{rel}\sigma(k \cdot u_{\Upsilon}) \rangle =  \frac{\int d^{3}k~v_{rel}~\sigma(k \cdot u_{\Upsilon})~g(k^{0},T)}
          {\int d^{3}k~g(k^{0},T)}
\end{equation}

where g$(k^{0},T)$ is the gluon distribution in the rest frame of the QGP medium and  $v_{rel}$ is
relative velocity between an $\Upsilon$ and a gluon. The gluon-$\Upsilon$ dissociation cross-section
is given by \cite{SuppVineetShukla}

\begin{equation}\label{Dis_CS}
  \sigma(q^{0})  = \frac{2\pi}{3} \left (\frac{32}{3} \right)^{2} \frac{1}{m_{Q}(\varepsilon_{0}m_{Q})^{1/2}}
\frac{(q^{0}/\varepsilon_{0} - 1)^{3/2}}{(q^{0}/\varepsilon_{0})^{5}}
\end{equation}

where $m_{Q}$ is the $b$ quark mass and $q^{0}$ the gluon energy in the  rest frame of bottomonium and
its value must be larger than the binding energies ($\varepsilon_{0}$) of bottomonium states, by the
condition $(p + k)^2 \ge 4m_b^2$, implying that $\varepsilon = \varepsilon_{0} + (\varepsilon_{0})^2/(2m_{\Upsilon})$.
Now, the survival probability of the $\Upsilon$ states in such a dissociation reaction, averaged over the medium
boundary  is

\begin{equation}\label{Surv_diss}
  S_{diss}(R(N_{\rm part})) = \frac{\int_0^{R(N_{\rm part},\tau)}dr~r~exp^{-\int_{\tau_{0}}^{\tau_{QGP}} \Gamma_{D}n_{g}d\tau}}{\int_0^{R(N_{\rm part},\tau)}dr~r}.
\end{equation}

Where $\tau_{QGP}$ denotes the duration of QGP medium and the bottomonia with low momenta, in the medium of not
too high initial temperature, will in general stop interacting when the medium has cooled down enough.
The dissociation rates of bottomonium states are shown in Fig.~\ref{fig:Diss_Rate} (Left). $\Upsilon$ states like
$\Upsilon$(1S) with higher binding energy exhibit higher survival rate as compared to other excited states as
reflected in the final results of $R_{AA}$. Also, the states with larger transverse momenta is heavily affected by the
gluonic dissociation reaction. But this is recompensed in the color-screening reaction where higher p$_T$ bound states
are easily survived as shown in Ref.~\cite{URatioAbd}.

The fact that the recombination rate is depending on the dissociation cross-section which
greatly relies on the binding energy of the bottomonium bound states, makes the recombination process more appealing.
Also the factors such as the higher QGP-lifetime at LHC energy ($\approx$8-10 fm) and the bulk production of heavy
quarks ($c$, $b$) are playing significant role in the recombination process.
Prior to moving towards the solution of recombination rate, the probability that some of the newly formed $\Upsilon$
may dissociate with thermal gluons (even though the rate will be very low in the beginning but the number could be
significant at later stages of QGP evolution), is to be taken into account in the calculation of total survival rates.
Eq.~(\ref{Decoup_Reco1}) then becomes

\begin{equation}\label{Decoup_Reco2}
\frac{dN_{\Upsilon}^{F}}{d\tau} = \Gamma_{F}N_{b\overline{b}}^{2}(Tot)[V(\tau)]^{-1} - \Gamma_{D}N^{F}_{\Upsilon}n_{g}.
\end{equation}

Where $N_{b\overline{b}}(Tot)$ is the sum of the $b\overline{b}$ pair produced initially in the collisions, $N_{b\overline{b}}(0)$ and
those $b$ and $\overline{b}$ quarks separated due to the dissociation of bottomonium states, $N_{b\overline{b}}^{Diss}$. 
The $N_{b\overline{b}}(Tot)$ = $ N_{b\overline{b}}(0) + N_{b\overline{b}}^{Diss}$. The second term in the RHS of the
Eq.~(\ref{Decoup_Reco2}) is for the assessment of the dissociation of newly made $\Upsilon$s.

\renewcommand{\arraystretch}{1.5}
\begin{table*}
  \begin{center}
    \caption{\label{DecayReco} Solution of the differential equation of recombination process (Eq.~(\ref{Decoup_Reco2}))
      using Bateman equation which is used commonly for the solution of Nuclear decay chain reactions.}
\begin{tabular}{p{5cm}|p{6cm}}
\hline
  {\rm Nuclear decay process}        & Recombination process    \\
\hline
$\frac {dN_1} {dt}  = - \lambda_1 N_1 $   with  $N_1 = N^0_1 e^{\lambda_1 t}$     & $\frac {dN^F_{\Upsilon}} {dt}  = - \lambda_{b\bar b}N_{b\bar b}$ (creation from b \& $\bar b$)   \\
$\frac {dN_2} {dt}  = - \lambda_2 N_2  +  \lambda_1 N_1 $ &
$\frac {dN^F_{\Upsilon}}{dt}  = \lambda_{b \bar b} N_{b \bar b} - \lambda_{D} N^F_{\Upsilon} $ \\
 (Daughter nuclei decay) & (Recombination \& Dissociation of newly created $\Upsilon$) \\
$N_2 = \frac{\lambda_1}{\lambda_2 - \lambda_1} N_0 (e^{\lambda_1 t} - e^{\lambda_1 t})  + N_2 e^{\lambda_1 t}$ (Bateman Solution) &
$N_{\Upsilon}^{F}$ is calculated in Eq.~(\ref{Decoup_RecoTot}) \\
\hline
    \end{tabular}
\end{center}
\end{table*}

The new formation rate, Eq.~(\ref{Decoup_Reco2}) looks like a radioactive decay chain reaction.
In the decay chain, the parent nucleus decays to daughter nuclei (here in place of decay,
$\Upsilon$ is made by the recombination of the $b$ and $\overline{b}$ quarks and the number of bottom quarks
go down as the formation rate goes up). After that daughter nuclei decays once again to third nuclei
(here newly formed $\Upsilon$s dissociate). Likewise, the recombination rate including dissociation of
newly created $\Upsilon$
states is given in the form of an equation $\frac {dN^F_{\Upsilon}}{dt}  = \lambda_{b \bar b} N_{b \bar b} - \lambda_{D} N^F_{\Upsilon} $
as depicted in Table~\ref{DecayReco}.
This is similar to Eq.~(\ref{Decoup_Reco2}) and the solution of such differential equation can be
achieved by the Bateman equation which incorporates the effects of correlated mechanisms
of recombination from two $b$ quarks and the dissociation of newly created $\Upsilon$ states~\cite{Bateman}.
The solution, a first of its kind taken in this paper, becomes as follows

\begin{align}\label{Decoup_RecoTot}
 N_{\Upsilon}^{F} & = \frac{\Lambda_{F}} {\Lambda_{D} - \Lambda_{F}}~N_{b\overline{b}}(Tot)
  [ e^{-\int_{\tau_{0}}^{\tau_{QGP}} \Gamma_{F}N_{b\overline{b}}^{2}(Tot)[V(\tau)]^{-1}d\tau}
      - e^{-\int_{\tau_{0}}^{\tau_{QGP}} \Gamma_{D}n_gd\tau} ] \nonumber \\
   & +~ N_{b\overline{b}}^{Diss}~e^{-\int_{\tau_{0}}^{\tau_{QGP}} \Gamma_{D}n_gd\tau}   
\end{align}
with $\Lambda_{F}$  =  $\int_{\tau_{0}}^{\tau_{QGP}} \Gamma_{F}N_{b\overline{b}}^{2}(Tot)[V(\tau)]^{-1}d\tau$ and
$\Lambda_{D}$  = $\int_{\tau_{0}}^{\tau_{QGP}} \Gamma_{D}n_gd\tau$.

 The probability of recombination (fractional of the formation/recombination) in the medium becomes

\begin{align}\label{Decoup_RecoFrac}
  S_F(N_{\rm part}) & = \frac{N_{\Upsilon}^{F}}{N_{\Upsilon}(0) + N_{b\overline{b}}(Tot)} =  \notag \\
  & \frac{N_{b\overline{b}}(Tot)}{N_{\Upsilon}(0) + N_{b\overline{b}}(Tot)} \frac{\Lambda_{F}} {\Lambda_{D} - \Lambda_{F}} 
  [ e^{-\int_{\tau_{0}}^{\tau_{QGP}} \Gamma_{F}N_{b\overline{b}}^{2}(Tot)[V(\tau)]^{-1}d\tau} - \notag\\ 
 &   e^{-\int_{\tau_{0}}^{\tau_{QGP}} \Gamma_{D}n_gd\tau} ]
  + ~\frac{N_{b\overline{b}}^{Diss}}{N_{\Upsilon}(0) + N_{b\overline{b}}(Tot)}~e^{-\int_{\tau_{0}}^{\tau_{QGP}} \Gamma_{D}n_gd\tau}.
\end{align}

The net survival probability of the bottomonium in the medium is the combined effects from
all interactions, Eq.~(\ref{Surv_Color}), Eq.(\ref{Surv_diss}) and Eq.(\ref{Decoup_RecoFrac}).

\begin{align}\label{Surv_Tot}
  S(N_{part}) = S_{col}(N_{\rm part}) * S_{diss}(N_{\rm part}) + S_F(N_{\rm part}).
\end{align}
  
The Eq.~(\ref{Surv_Tot}) is the nuclear suppression factor, $R_{AA}$ and is obtained by comprising
the feed-down corrections~\cite{Initials1,Initials2}. 

\section{Results and discussions}

\begin{figure}[htb]
\begin{center}
\begin{tabular}{cc}
\includegraphics[width=0.48\textwidth]{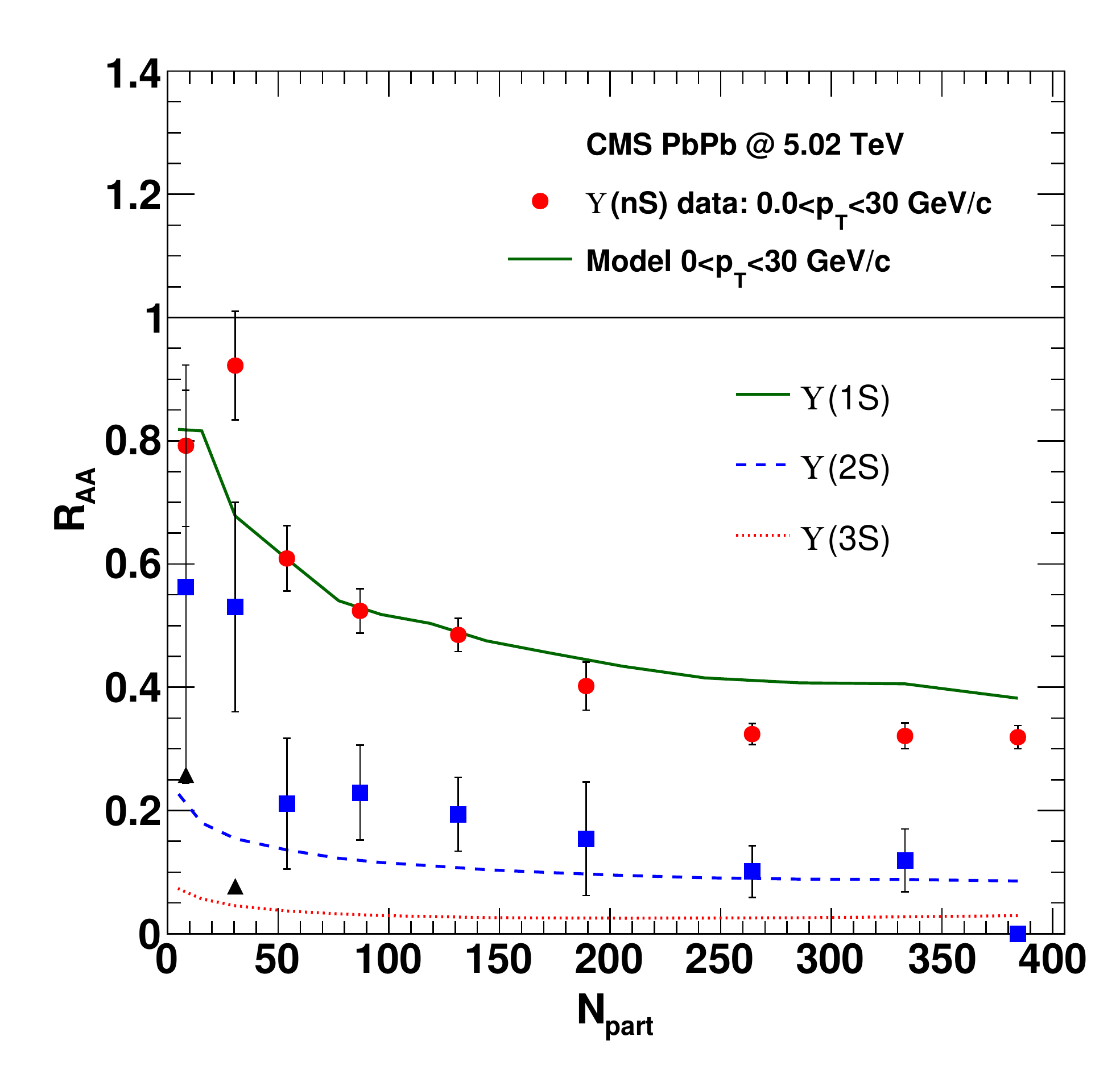}&
\includegraphics[width=0.49\textwidth]{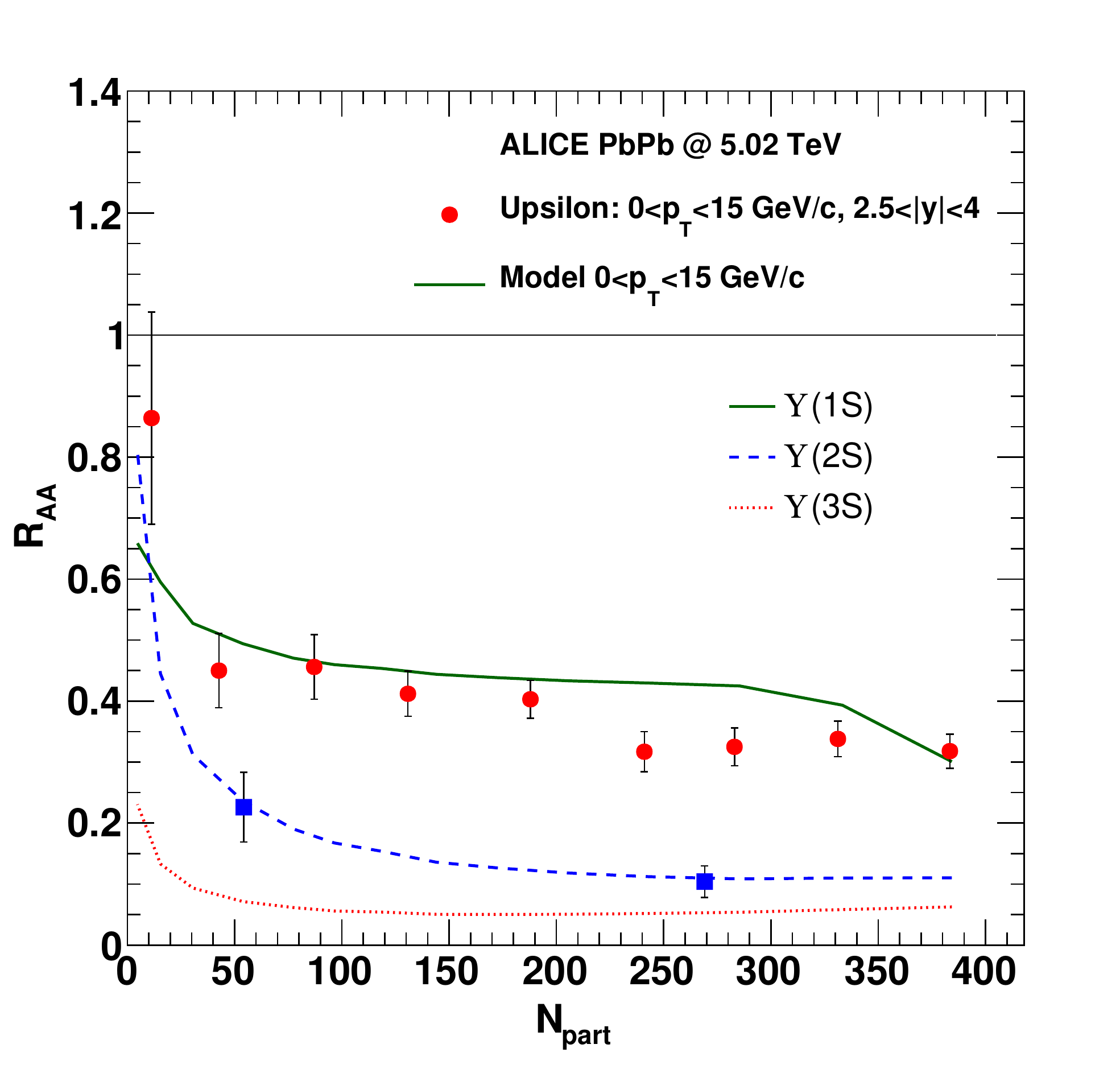}\\
\end{tabular}
\caption{The nuclear modification factor, $R_{AA}$ as a function of $N_{\rm part}$ for three $\Upsilon$(nS)
  states. The solid points are measured $R_{AA}$ by CMS experiment in Pb+Pb collisions at $\sqrt{s_{\rm NN}}$
  = 5.02 TeV (Left) and the $R_{AA}$ measured with ALICE experiment at low p$_T$ and forward rapidity (Right).
  The solid and dashed lines in all figures represent the model calculations.}
\label{fig:UpsiRaaCMSALICE}
\end{center}
\end{figure}

With present model of decoupled rates of dissociation and recombination, we computed the nuclear modification
factors $R_{AA}$ of three $\Upsilon$(nS) states as function of number of participants $N_{\rm part}$ and transverse momentum $p_{T}$. The model calculations are compared with the data measured at CMS and ALICE experiments. As mentioned above,
many parameters are inter-playing in the determination of the final trend of the bottomonium suppression.
The survival probabilities
of resonance states have a special $p_T$ reliance determined by the $T_D$, $\tau_{F}$, $T(\tau)$ and $\tau_{med}$ for
every bottomonium state. The $R_{AA}$ of $\Upsilon$(nS) as a function of $N_{\rm part}$ is plotted in Fig.~\ref{fig:UpsiRaaCMSALICE}.
The solid circles, squares and triangles are the measured $R_{AA}$ for $\Upsilon$(1S), $\Upsilon$(2S) and
$\Upsilon$(3S) respectively with CMS experiment (Left) and ALICE experiment (Right) in Pb+Pb collisions
at $\sqrt{s_{\rm NN}}$ = 5.02 TeV \cite{CMS502TeV_UpsiRaa}. The higher value of binding energy and smaller $\tau_{F}$ 
help the ground state $\Upsilon$(1S) get relatively less suppressed and smaller binding energy and higher value of $\tau_{F}$ of
the excited states make them more vulnerable during the longer duration of QGP. The predicted scenario of sequential
suppression of bottomonia is validated in both calculations. Although the model calculations recreate the
trends of the suppression very nicely, there appears an incompatibility in some kinematic regions.

\begin{figure}[htb]
\begin{center}
\includegraphics[width=0.48\textwidth]{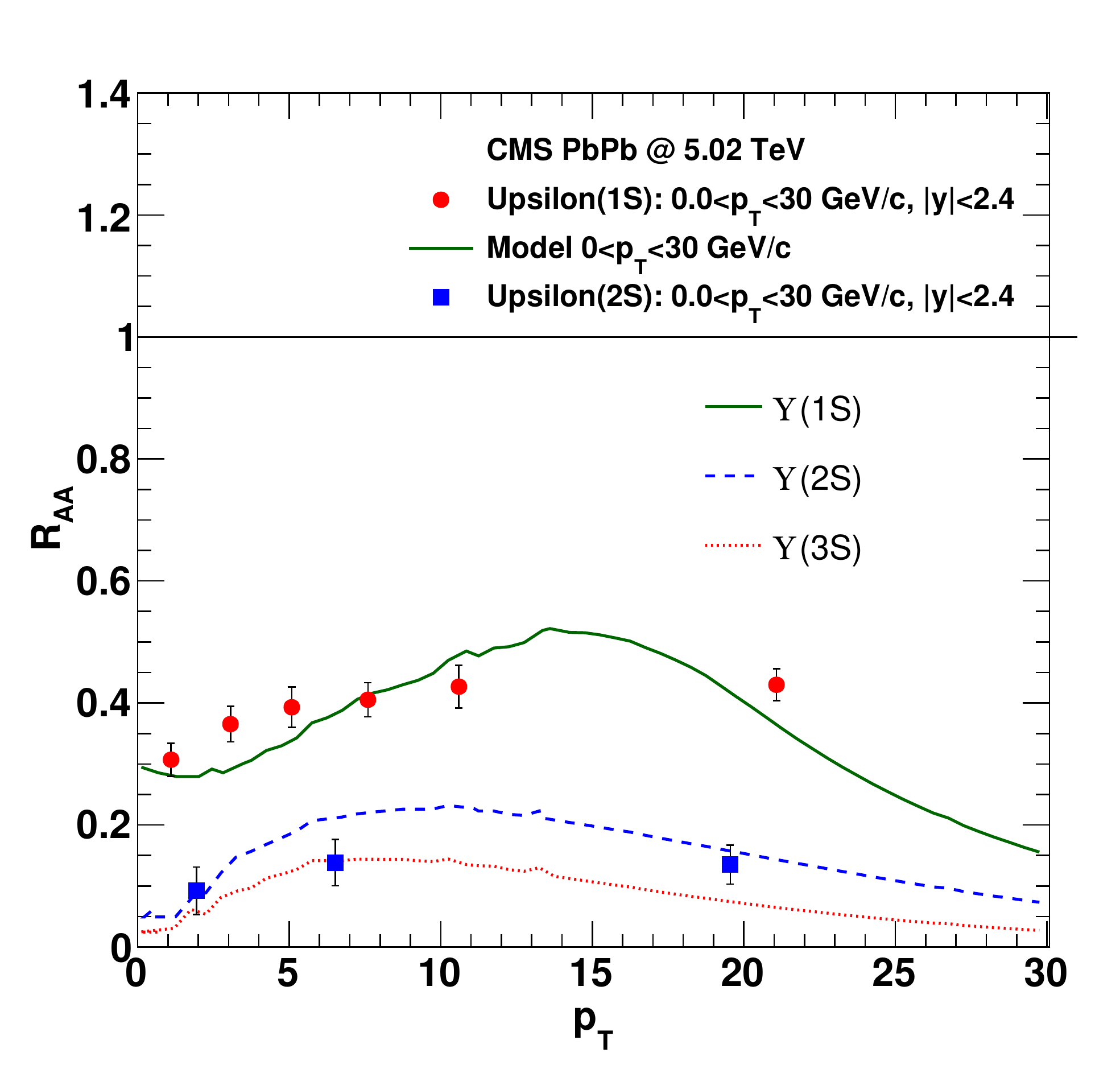} 
\caption{The nuclear modification factor, $R_{AA}$ as a function of $p_{T}$ (0-30 GeV/c) for three
  $\Upsilon$(nS) states. The solid points are the measured $R_{AA}$ by CMS experiment in Pb+Pb collisions
  at $\sqrt{s_{\rm NN}}$ = 5.02 TeV. The lines represent the present model calculations.} 
\label{fig:UpsiRaaPt}
\end{center}
\end{figure}

\begin{figure}[htb]
\begin{center}
\includegraphics[width=0.48\textwidth]{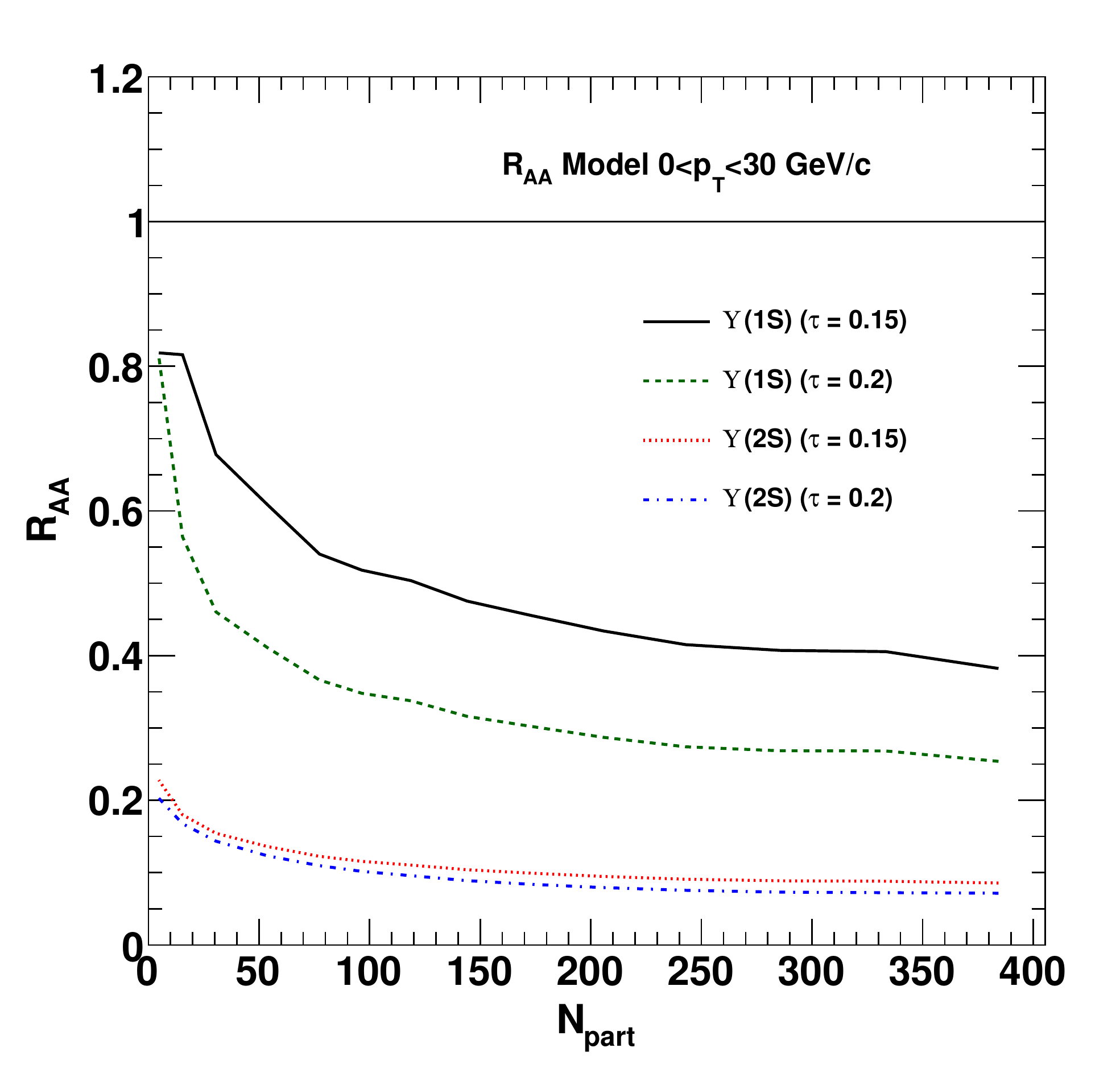} 
\caption{The nuclear modification factor, $R_{AA}$ as a function of $N_{\rm part}$ for $\Upsilon$(1S) and
  $\Upsilon$(2S) states with different initial values for $\tau_{0}$.}
\label{fig:UpsiRaaTau}
\end{center}
\end{figure}

Likewise, the $R_{AA}$ of three states as a function of $p_{T}$ (0-30 GeV/c) is depicted in Fig.~\ref{fig:UpsiRaaPt}.
The dashed and solid lines are showing the model calculations for $R_{AA}$ in the relevant $p_{T}$ bins.
The interaction between various medium-induced reactions determines the pattern of $p_{T}$ curve in every region.
The model calculations reproduce the patterns of the observed $R_{\rm AA}$ in majority of the $p_{T}$ region.
There is a slight tendency of less suppression as we go from low $p_T$ bins to high $p_T$ bins.
This might be due to lower energy loss of high $p_{T}$ bottomonia as anticipated in the energy loss model~\cite{Fransua}.
Also, the variation of initial parameters such as $T_0$ and $\tau_{0}$ which are out of experimental measurements
gives rise to finite modification in the suppression pattern of the model calculations. Since the $\tau_{0}$ value
modulates many factors like dissociation time, screening region and etc., its variation can cause a significant
impact on the modification of $\Upsilon$(nS) states. The Fig.~\ref{fig:UpsiRaaTau} shows such an impact on the
suppression curves of $\Upsilon$(1S) and $\Upsilon$(2S) states due to the changes in $\tau_{0}$ values. 

\section{Conclusions}
In this article, the nuclear modification factors of bottomonia states ($\Upsilon$(nS)) in an inflating QGP
created in Pb+Pb collisions at $\sqrt{s_{NN}}=$ 5.02 TeV have been evaluated
using a framework of decoupled rates of dissociation and recombination. The nuclear modification comes from
the net effect of color screening together
with gluon-dissociation and recombination reactions. The tussle between the time $\tau_{0}$, $\tau_{F}$, the temperature $T(\tau)$,
binding energy, etc. determine the epic of the survival probabilities of $\Upsilon$ states in various kinematic regions.
The evaluated suppressions are matched with the $R_{AA}$ observed at CMS and ALICE Experiments in the kinematic bins of
centrality and $p_{T}$. The model calculation recreates nicely the measured $R_{AA}$ of $\Upsilon$ states.
Moreover, the trend of sequential suppression in the observed $R_{AA}$ at LHC experiments has been manifested
nicely in the model calculations. Besides the modification from the medium reactions, variation of the initial
parameters can cause a significant impact on the suppression pattern of $\Upsilon$(nS) states.



\end{document}